\pdfoutput=1

\documentclass[12pt,a4paper]{article}

\usepackage{ifthen} 
\newboolean{pdflatex}
\setboolean{pdflatex}{true} 

\newboolean{articletitles}
\setboolean{articletitles}{true} 

\newboolean{uprightparticles}
\setboolean{uprightparticles}{false} 

\newboolean{inbibliography}
\setboolean{inbibliography}{false} 

\usepackage[top=1in, bottom=1.25in, left=1in, right=1in]{geometry}

\usepackage{microtype}
\usepackage{lineno}  
\usepackage{xspace} 
\usepackage{caption} 

\usepackage{graphicx}  
\usepackage{color}
\usepackage{colortbl}
\graphicspath{{./figs/}} 

\usepackage{amsmath} 
\usepackage{amssymb}
\usepackage{amsfonts}
\usepackage{upgreek} 

\newcommand*\patchAmsMathEnvironmentForLineno[1]{%
\expandafter\let\csname old#1\expandafter\endcsname\csname #1\endcsname
\expandafter\let\csname oldend#1\expandafter\endcsname\csname
end#1\endcsname
 \renewenvironment{#1}%
   {\linenomath\csname old#1\endcsname}%
   {\csname oldend#1\endcsname\endlinenomath}%
}
\newcommand*\patchBothAmsMathEnvironmentsForLineno[1]{%
  \patchAmsMathEnvironmentForLineno{#1}%
  \patchAmsMathEnvironmentForLineno{#1*}%
}
\AtBeginDocument{%
\patchBothAmsMathEnvironmentsForLineno{equation}%
\patchBothAmsMathEnvironmentsForLineno{align}%
\patchBothAmsMathEnvironmentsForLineno{flalign}%
\patchBothAmsMathEnvironmentsForLineno{alignat}%
\patchBothAmsMathEnvironmentsForLineno{gather}%
\patchBothAmsMathEnvironmentsForLineno{multline}%
\patchBothAmsMathEnvironmentsForLineno{eqnarray}%
}

\usepackage{hyperref}    
\usepackage[all]{hypcap} 

\usepackage{xspace} 
\usepackage{upgreek}

\def\lhcb {\mbox{LHCb}\xspace}

\def\babar  {\mbox{BaBar}\xspace}
\def\belle  {\mbox{Belle}\xspace}
\def\cleo   {\mbox{CLEO}\xspace}

\def\MagUp {\mbox{\em Mag\kern -0.05em Up}\xspace}

\ifthenelse{\boolean{uprightparticles}}%
{

 \def\Ppi         {\ensuremath{\uppi}\xspace}                 
                  
 \def\Prho        {\ensuremath{\uprho}\xspace}

 \def\PDelta      {\ensuremath{\Delta}\xspace}                 
 \def\PXi      {\ensuremath{\Xi}\xspace}                 
 \def\PLambda      {\ensuremath{\Lambda}\xspace}                 
 \def\PSigma      {\ensuremath{\Sigma}\xspace}                 
 \def\POmega      {\ensuremath{\Omega}\xspace}                 
 \def\PUpsilon      {\ensuremath{\Upsilon}\xspace}                 
 
 \def\PB      {\ensuremath{\mathrm{B}}\xspace}                 
                  
 \def\PD      {\ensuremath{\mathrm{D}}\xspace}

 \def\PK      {\ensuremath{\mathrm{K}}\xspace}

 \def\Pb      {\ensuremath{\mathrm{b}}\xspace}                 
 \def\Pc      {\ensuremath{\mathrm{c}}\xspace}

 \def\Pi      {\ensuremath{\mathrm{i}}\xspace}

 \def\Pp      {\ensuremath{\mathrm{p}}\xspace}

 \def\Ps      {\ensuremath{\mathrm{s}}\xspace}

}
{

 \def\Ppi         {\ensuremath{\pi}\xspace}                 
                  
 \def\Prho        {\ensuremath{\rho}\xspace}

 \mathchardef\PDelta="7101
 \mathchardef\PXi="7104
 \mathchardef\PLambda="7103
 \mathchardef\PSigma="7106
 \mathchardef\POmega="710A
 \mathchardef\PUpsilon="7107
                  
 \def\PB      {\ensuremath{B}\xspace}                 
                  
 \def\PD      {\ensuremath{D}\xspace}

 \def\PK      {\ensuremath{K}\xspace}

 \def\Pb      {\ensuremath{b}\xspace}                 
 \def\Pc      {\ensuremath{c}\xspace}

 \def\Pi      {\ensuremath{i}\xspace}

 \def\Pp      {\ensuremath{p}\xspace}

 \def\Ps      {\ensuremath{s}\xspace}

}

\makeatletter
\ifcase \@ptsize \relax
  \newcommand{\miniscule}{\@setfontsize\miniscule{4}{5}}
\or
  \newcommand{\miniscule}{\@setfontsize\miniscule{5}{6}}
\or
  \newcommand{\miniscule}{\@setfontsize\miniscule{5}{6}}
\fi
\makeatother

\DeclareRobustCommand{\optbar}[1]{\shortstack{{\miniscule (\rule[.5ex]{1.25em}{.18mm})}
  \\ [-.7ex] $#1$}}

\def\squark    {{\ensuremath{\Ps}}\xspace}

\def\cquark    {{\ensuremath{\Pc}}\xspace}

\def\bquark    {{\ensuremath{\Pb}}\xspace}

\def\pion   {{\ensuremath{\Ppi}}\xspace}

\def\pip    {{\ensuremath{\pion^+}}\xspace}
\def\pim    {{\ensuremath{\pion^-}}\xspace}

\def\rhomeson {{\ensuremath{\Prho}}\xspace}

\def\rhom     {{\ensuremath{\rhomeson^-}}\xspace}

\def\kaon    {{\ensuremath{\PK}}\xspace}
  \def\Kbar    {{\kern 0.2em\overline{\kern -0.2em \PK}{}}\xspace}

\def\KorKbar    {\kern 0.18em\optbar{\kern -0.18em K}{}\xspace}

\def\Kp      {{\ensuremath{\kaon^+}}\xspace}
\def\Km      {{\ensuremath{\kaon^-}}\xspace}

\def\KS      {{\ensuremath{\kaon^0_{\mathrm{ \scriptscriptstyle S}}}}\xspace}

\def\Kstarm  {{\ensuremath{\kaon^{*-}}}\xspace}

  \def\Dbar    {{\kern 0.2em\overline{\kern -0.2em \PD}{}}\xspace}
\def\D       {{\ensuremath{\PD}}\xspace}

\def\DorDbar    {\kern 0.18em\optbar{\kern -0.18em D}{}\xspace}
\def\Dz      {{\ensuremath{\D^0}}\xspace}

\def\B       {{\ensuremath{\PB}}\xspace}
\def\Bbar    {{\ensuremath{\kern 0.18em\overline{\kern -0.18em \PB}{}}}\xspace}

\def\BorBbar    {\kern 0.18em\optbar{\kern -0.18em B}{}\xspace}

\def\Bu      {{\ensuremath{\B^+}}\xspace}

\def\Bd      {{\ensuremath{\B^0}}\xspace}
\def\Bs      {{\ensuremath{\B^0_\squark}}\xspace}
\def\Bsb     {{\ensuremath{\Bbar{}^0_\squark}}\xspace}

\def\Bc      {{\ensuremath{\B_\cquark^+}}\xspace}

  \def\Y#1S{\ensuremath{\PUpsilon{(#1S)}}\xspace}

\def\proton      {{\ensuremath{\Pp}}\xspace}
\def\antiproton  {{\ensuremath{\overline \proton}}\xspace}

\def\Lz          {{\ensuremath{\PLambda}}\xspace}
\def\Lbar        {{\ensuremath{\kern 0.1em\overline{\kern -0.1em\PLambda}}}\xspace}
\def\LorLbar    {\kern 0.18em\optbar{\kern -0.18em \PLambda}{}\xspace}

\def\Sigmares    {{\ensuremath{\PSigma}}\xspace}
\def\Sigmaresbar {{\ensuremath{\overline \Sigmares}}\xspace}
\def\Sigmaz      {{\ensuremath{\Sigmares^0}}\xspace}
\def\Sigmazbar   {{\ensuremath{\Sigmaresbar{}^0}}\xspace}

\def\Lb      {{\ensuremath{\Lz^0_\bquark}}\xspace}

\def\Lc      {{\ensuremath{\Lz^+_\cquark}}\xspace}
\def\Lcbar   {{\ensuremath{\Lbar{}^-_\cquark}}\xspace}

\def\BF         {{\ensuremath{\mathcal{B}}}\xspace}

\def\BR         {\BF}
\newcommand{\decay}[2]{\ensuremath{#1\!\to #2}\xspace}         

\def\to                 {\ensuremath{\rightarrow}\xspace}

\def\CP                {{\ensuremath{C\!P}}\xspace}

\def\AT#1     {\ensuremath{A_{\mathrm{T}}^{#1}}\xspace}           

\def\C#1      {\ensuremath{\mathcal{C}_{#1}}\xspace}                       
\def\Cp#1     {\ensuremath{\mathcal{C}_{#1}^{'}}\xspace}                    
\def\Ceff#1   {\ensuremath{\mathcal{C}_{#1}^{\mathrm{(eff)}}}\xspace}        
\def\Cpeff#1  {\ensuremath{\mathcal{C}_{#1}^{'\mathrm{(eff)}}}\xspace}       
\def\Ope#1    {\ensuremath{\mathcal{O}_{#1}}\xspace}                       
\def\Opep#1   {\ensuremath{\mathcal{O}_{#1}^{'}}\xspace}                    

\newcommand{\tev}{\ifthenelse{\boolean{inbibliography}}{\ensuremath{~T\kern -0.05em eV}}{\ensuremath{\mathrm{\,Te\kern -0.1em V}}}\xspace}
\newcommand{\gev}{\ensuremath{\mathrm{\,Ge\kern -0.1em V}}\xspace}
\newcommand{\mev}{\ensuremath{\mathrm{\,Me\kern -0.1em V}}\xspace}
\newcommand{\kev}{\ensuremath{\mathrm{\,ke\kern -0.1em V}}\xspace}
\newcommand{\ev}{\ensuremath{\mathrm{\,e\kern -0.1em V}}\xspace}
\newcommand{\gevc}{\ensuremath{{\mathrm{\,Ge\kern -0.1em V\!/}c}}\xspace}
\newcommand{\mevc}{\ensuremath{{\mathrm{\,Me\kern -0.1em V\!/}c}}\xspace}
\newcommand{\gevcc}{\ensuremath{{\mathrm{\,Ge\kern -0.1em V\!/}c^2}}\xspace}
\newcommand{\gevgevcccc}{\ensuremath{{\mathrm{\,Ge\kern -0.1em V^2\!/}c^4}}\xspace}
\newcommand{\mevcc}{\ensuremath{{\mathrm{\,Me\kern -0.1em V\!/}c^2}}\xspace}

\def\invfb   {\ensuremath{\mbox{\,fb}^{-1}}\xspace}

\newcommand{\chisq}{\ensuremath{\chi^2}\xspace}

\newcommand{\chisqip}{\ensuremath{\chi^2_{\text{IP}}}\xspace}

\def\gsim{{~\raise.15em\hbox{$>$}\kern-.85em
          \lower.35em\hbox{$\sim$}~}\xspace}
\def\lsim{{~\raise.15em\hbox{$<$}\kern-.85em
          \lower.35em\hbox{$\sim$}~}\xspace}

\def\sPlot{\mbox{\textit{sPlot}}\xspace}

\def\ptot       {\mbox{$p$}\xspace}
\def\pt         {\mbox{$p_{\mathrm{ T}}$}\xspace}

\def\tell1  {TELL1\xspace}
\def\ukl1   {UKL1\xspace}

\newcommand{\ie}{\mbox{\itshape i.e.}\xspace}

\def\Bds   {\mbox{$\B^0_{(\squark)}$}\xspace}

\def\BPLbarH  {\ensuremath{\Bds \to \proton \Lbar h^-}\xspace}

\def\PLbar  {\ensuremath{\proton \Lbar}\xspace}

\def\Sbar{\Sigmazbar}

\def\PLbarH {\ensuremath{\proton \Lbar h^-}\xspace}
\def\PLbarPi{\ensuremath{\proton \Lbar \pim}\xspace}
\def\PLbarK {\ensuremath{\proton \Lbar \Km}\xspace}

\def\BPSbarH  {\ensuremath{\Bds \to \proton \Sbar h^-}\xspace}

\newcommand{\LPPi}{\texorpdfstring{\decay{\PLambda}{\proton \pim}}{}}

\newcommand{\LbLHH}{\texorpdfstring{\decay{\Lb}{\Lz h^{+} h^{\prime-}}}{}}
\newcommand{\LbLPPbar}{\texorpdfstring{\decay{\Lb}{\Lz \proton \antiproton}}{}}

\newcommand{\LcPKPi}{\texorpdfstring{\decay{\Lc}{\proton \Km \pip}}{}}

\newcommand{\LcbarLbarPi}{\texorpdfstring{\decay{\Lcbar}{\Lbar \pim}}{}}
\newcommand{\LcbarLbarK}{\texorpdfstring{\decay{\Lcbar}{\Lbar \Km}}{}}

\newcommand{\BdPLbarPi}{\texorpdfstring{\decay{\Bd}{\proton \Lbar \pim}}{}}
\newcommand{\BdPLbarK}{\texorpdfstring{\decay{\Bd}{\proton \Lbar \Km}}{}}
\newcommand{\BsPLbarK}{\texorpdfstring{\decay{\Bs}{\proton \Lbar \Km}}{}}
\newcommand{\BsPbarLK}{\texorpdfstring{\decay{\Bs}{\antiproton \Lz \Kp}}{}}

\newcommand{\BsPLbarPi}{\texorpdfstring{\decay{\Bs}{\proton \Lbar \pim}}{}}

\newcommand{\BdPSbarPi}{\texorpdfstring{\decay{\Bd}{\proton \Sbar \pim}}{}}
\newcommand{\BsPSbarK}{\texorpdfstring{\decay{\Bs}{\proton \Sbar \Km}}{}}

\newcommand{\BdPLbarRho}{\texorpdfstring{\decay{\Bd}{\proton \Lbar \rhom}}{}}

\newcommand{\BsPLbarKstar}{\texorpdfstring{\decay{\Bs}{\proton \Lbar \Kstarm}}{}}

\newcommand{\BsLcbarLPi}{\texorpdfstring{\decay{\Bs}{\Lcbar \Lz \pip}}{}}
\newcommand{\BdLcbarP}{\texorpdfstring{\decay{\Bd}{\Lcbar \proton}}{}}

\def\c     {\ensuremath{c}\xspace}

\newcommand{\KSPiPi}{\texorpdfstring{\decay{\KS}{\pip \pim}}{}}

\usepackage{cite} 
\usepackage{mciteplus}

\usepackage{longtable} 

\begin{document}

\renewcommand{\thefootnote}{\fnsymbol{footnote}}
\setcounter{footnote}{1}

\begin{titlepage}
\pagenumbering{roman}

\vspace*{-1.5cm}
\centerline{\large EUROPEAN ORGANIZATION FOR NUCLEAR RESEARCH (CERN)}
\vspace*{1.5cm}
\noindent
\begin{tabular*}{\linewidth}{lc@{\extracolsep{\fill}}r@{\extracolsep{0pt}}}
\ifthenelse{\boolean{pdflatex}}
{\vspace*{-2.7cm}\mbox{\!\!\!\includegraphics[width=.14\textwidth]{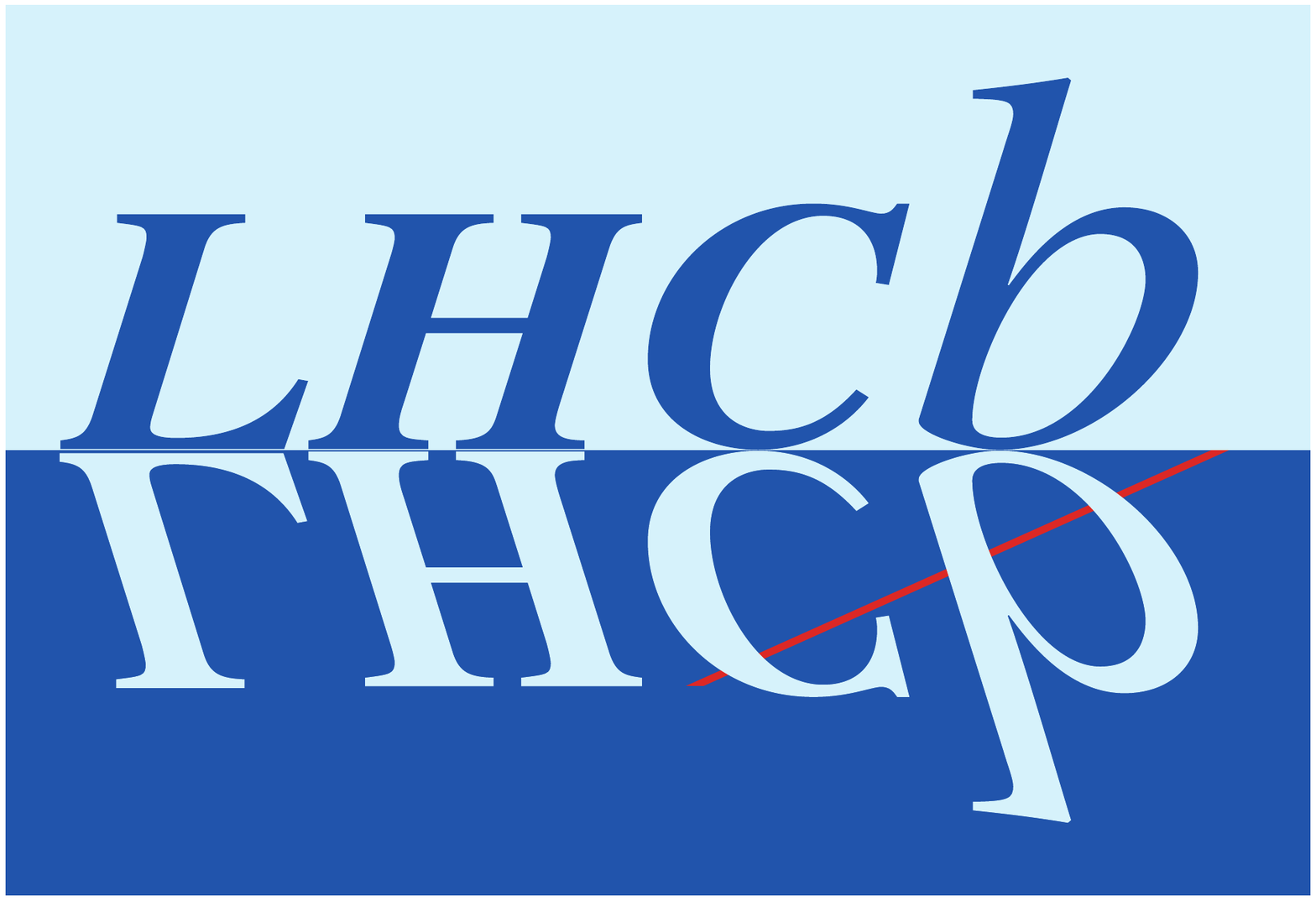}} & &}%
{\vspace*{-1.2cm}\mbox{\!\!\!\includegraphics[width=.12\textwidth]{lhcb-logo.eps}} & &}%
\\
 & & CERN-EP-2017-067 \\  
 & & LHCb-PAPER-2017-012 \\  
 & & 25 July 2017 \\ 
 & & \\
\end{tabular*}

\vspace*{4.0cm}

{\normalfont\bfseries\boldmath\huge
\begin{center}
First observation\\of a baryonic \Bs decay
\end{center}
}

\vspace*{2.0cm}

\begin{center}
The LHCb collaboration\footnote{Authors are listed at the end of this paper.}
\end{center}

\vspace{\fill}

\begin{abstract}
  \noindent
We report the first observation of a baryonic \Bs decay, \BsPLbarK,
using proton-proton collision data recorded by the \lhcb experiment at
center-of-mass energies of 7 and 8\tev,
corresponding to an integrated luminosity of 3.0\invfb.
The branching fraction is measured to be
$
\BF(\BsPLbarK) + \BF(\BsPbarLK) = \left[\vphantom{\sum}{5.46} \pm{0.61} \pm 0.57 \pm 0.50 (\BR) \pm 0.32 (f_s/f_d)\right]\times 10^{-6},
$
where the first uncertainty is statistical and the second systematic,
the third uncertainty accounts for the experimental uncertainty on the
branching fraction of the \BdPLbarPi decay used for normalization,
and the fourth uncertainty relates to the knowledge of the
ratio of $b$-quark hadronization probabilities $f_s/f_d$.
\end{abstract}

\vspace*{2.0cm}

\begin{center}
  Published in Phys.~Rev.~Lett. 119 (2017) 041802
\end{center}

\vspace{\fill}

{\footnotesize 
\centerline{\copyright~CERN on behalf of the \lhcb collaboration, licence \href{http://creativecommons.org/licenses/by/4.0/}{CC-BY-4.0}.}}
\vspace*{2mm}

\end{titlepage}


\newpage
\setcounter{page}{2}
\mbox{~}
%

\cleardoublepage


\renewcommand{\thefootnote}{\arabic{footnote}}
\setcounter{footnote}{0}



\pagestyle{plain} 
\setcounter{page}{1}
\pagenumbering{arabic}


%

\noindent
The experimental study of $B$-meson decays to baryonic final states has a long history,
starting with the first observation of baryonic $B$ decays by the \cleo collaboration
in 1997~\cite{PhysRevLett.79.3125}.
The asymmetric $e^+ e^-$ collider experiments \babar and \belle reported numerous searches
and observations of decays of \Bd and \Bu mesons to baryonic final states~\cite{Bevan:2014iga}.
The \lhcb collaboration published the first observation
of a baryonic \Bc decay in 2014~\cite{LHCb-PAPER-2014-039}.
Until now, no baryonic \Bs decay has ever been observed with a significance in excess of
five standard deviations;
the \belle collaboration provided the only evidence for such a process
in the study of \BsLcbarLPi decays,
with a significance of 4.4 standard deviations~\cite{Solovieva:2013rhq}.

Areas of particular interest in baryonic \B decays are the study of the
hierarchy of branching fractions and the threshold enhancement
in the baryon-antibaryon mass spectrum~\cite{Hou:2000bz,Bevan:2014iga}.
Multi-body baryonic \B decays are expected to have higher branching fractions
than two-body decays~\cite{Hsiao:2014zza,Cheng:2014qxa}.
The \BdPLbarPi and \BsPLbarK branching fractions are predicted to be of the order of
$10^{-6}$~\cite{Geng2017205}.
The notation \BsPLbarK is used hereafter for the sum of both accessible
final states $\BsPLbarK$ and $\BsPbarLK$.
As emphasized in Ref.~\cite{Geng2017205}, which studied the decays $\Bs \to \proton \Lbar h^-$,
the decay \BsPLbarK is a unique baryonic \B decay in that it is the only presently known decay where all four processes,
namely the decays of a \Bs or a \Bsb meson to either the $\proton \Lbar \Km$
or the $\antiproton \Lz \Kp$ final state, can occur.
A \B-flavor-tagged decay-time-dependent study is required in order to
separate the two possible final states and measure their individual branching fractions
as well as \CP violation observables.

The current experimental knowledge on the family of \BPLbarH decays (\mbox{$h = \pion, \kaon$})
and related modes such as \BPSbarH, with $\Sbar \to \Lbar \gamma$, is rather scarce.
The \BdPLbarPi decay has been studied by the
\babar~\cite{Aubert:2009am} and \belle~\cite{Wang:2003yi,Wang:2007as} collaborations and
the \belle collaboration has reported the 90\% confidence level upper limits
$\BF(\BdPLbarK) < 8.2 \times 10^{-7}$ and
$\BF(\BdPSbarPi) < 3.8 \times 10^{-6}$~\cite{Wang:2003yi}.

Manifestations of \CP and $T$ violation in baryonic $B$ decays have been studied
from a theoretical viewpoint, see for example Ref.~\cite{Geng:2008ps} and references therein.
A large \CP-violation asymmetry of order 10\% is expected for the \BdPLbarPi
decay mode~\cite{Geng:2008ps}, which further motivates the experimental study of
\BPLbarH decays.

This Letter presents the first observation of a charmless baryonic \Bs decay.
The branching fraction of the \BsPLbarK decay is measured relative to that
of the topologically identical \BdPLbarPi decay to suppress common systematic uncertainties:
\begin{equation}
\BF(\BsPLbarK) + \BF(\BsPbarLK) = \frac{f_d}{f_s}\,
                                  \frac{N(\BsPLbarK)}{N(\BdPLbarPi)}\,
                                  \frac{\epsilon_{\BdPLbarPi}}{\epsilon_{\BsPLbarK}}\,
                                  \BF(\BdPLbarPi) \,,
\label{eq:BF}
\end{equation}
\noindent
where $N$ represents yields determined from mass fits,
$f_q$ stands for the $\bquark$ hadronization probability to the meson $B_q$,
and $\epsilon$ represents the selection efficiencies.
The inclusion of charge-conjugate processes is implied, unless otherwise stated.

The data sample analyzed corresponds to an integrated luminosity of 1.0\invfb
of proton-proton collision data collected by the \lhcb experiment
at center-of-mass energies of 7\tev in 2011 and 2.0\invfb at 8\tev in 2012.
The \lhcb detector is a single-arm forward spectrometer covering the
\mbox{pseudorapidity} range $2 < \eta < 5$,
designed for the study of particles containing $b$ or $c$ quarks~\cite{Alves:2008zz,LHCb-DP-2014-002}.
The \mbox{pseudorapidity} is defined as $\eta=-{\rm ln}\left[\tan(\theta/2)\right]$,
where $\theta$ is the polar angle with respect to the proton in the positive $z$ direction.
The detector elements that are particularly
relevant to this analysis are a silicon-strip vertex detector surrounding the proton-proton interaction
region that allows heavy hadrons to be identified from their characteristically long
flight distance; a tracking system that provides a measurement of momentum, \ptot, of charged
particles; two ring-imaging Cherenkov detectors that are able to discriminate between
different species of charged hadrons; a calorimeter system
for the measurement of photons and neutral hadrons;
and multiwire proportional chambers for the detection of muons.
Simulated data samples, produced as described
in Refs.~\cite{Sjostrand:2007gs,*Sjostrand:2006za,LHCb-PROC-2010-056,Lange:2001uf,Golonka:2005pn,Allison:2006ve,*Agostinelli:2002hh,LHCb-PROC-2011-006},
are used to evaluate the response of the detector and to investigate and characterize
possible sources of background.

Events are selected in a similar way for
both the signal decay \BsPLbarK and the normalization channel \BdPLbarPi,
where $\Lbar\to\antiproton\pip$.
Real-time event selection is performed by a
trigger~\cite{LHCb-DP-2012-004} consisting of a hardware stage,
based on information from the calorimeter and muon systems, followed
by a software stage, which performs a full event reconstruction.
The hardware trigger stage requires events to have a muon with high
transverse momentum, \pt, or a
hadron, photon or electron with high transverse energy deposited in the calorimeters.
For this analysis, the hardware trigger decision can either be made on the signal candidates
or on other particles in the event.
The software trigger requires a two- or three-track
secondary vertex with a significant displacement from all the primary
$pp$ interaction vertices (PVs). At least one charged particle
must have high \pt and be inconsistent with originating from a PV.
A multivariate algorithm~\cite{BBDT} is used for
the identification of secondary vertices consistent with the decay
of a \bquark or \cquark hadron.

The \Lz decays are reconstructed in two different categories: the first consists of \Lz baryons
that decay early enough for the proton and pion to be
reconstructed in the vertex detector, while the second contains
those that decay later such that track segments cannot be reconstructed in
the vertex detector. These reconstruction categories are referred to as \emph{long}
and \emph{downstream}, respectively.

The selection of \Bds candidates, formed by combining
a \Lz candidate with a proton and a pion or kaon, is carried out with a
filtering stage, a requirement on the response of a
multilayer perceptron~\cite{MLPs} (MLP) classifier,
and particle identification (PID) criteria discussed below.
The proton and pion or kaon, of opposite charge, both decay products of the $B$ meson,
are hereafter referred to as the charged hadrons. Unless stated otherwise,
the terms proton and pion refer to the charged hadrons from the $B$-meson decay,
not to the \Lz decay products.
Both the \BdPLbarPi and the \BsPLbarK decay chains are
refitted~\cite{Hulsbergen:2005pu} employing a mass constraint on the \Lz candidates.

In the filtering stage the \Lz decay products are required to have a minimum momentum, \ptot,
form a good quality vertex
and satisfy $|m(\proton\pim) - m_\Lz| < 20 (15)\mevcc$ for downstream (long) candidates,
where $m_\Lz$ is the \Lz mass~\cite{PDG2016}.
They must have a large impact parameter (IP) with respect to all PVs,
where the IP is defined as the minimum distance of a track to a PV.
A minimum \chisqip with respect to any PV
is imposed on each \Lz decay product, where \chisqip is defined as the difference
between the vertex-fit \chisq of a PV reconstructed with and without the particle in question.
A loose PID requirement, based primarily on information
from the ring-imaging Cherenkov detectors, is imposed to select the proton candidate
from the \Lz baryon to remove background from \KS decays.
For downstream \Lz candidates a minimum momentum is also required.

A minimum requirement is imposed on the scalar sum of the \pt of the \Lz candidate
and the two charged hadrons.
The distance of closest approach among any pair from (p, \Lbar, $h^-$) divided by its uncertainty must be small.
The $B$ candidate must have a good quality vertex, have a minimum \pt and a small \chisqip
with respect to the associated PV as its reconstructed momentum vector should point to its production vertex;
the associated PV is the one with which it forms the smallest \chisqip.
The pointing condition of the $B$ candidate is further reinforced
by requiring that the angle between the \B-candidate momentum vector and the line
connecting the associated PV and the \B-decay vertex
($B$ direction angle, $\theta_B$) is close to zero.

Backgrounds from the \BdLcbarP decay with \LcbarLbarPi (\LcbarLbarK) are removed
from the \PLbarPi (\PLbarK) samples with a veto around the \Lc mass~\cite{PDG2016}
of three times the $\Lbar\pim$ ($\Lbar\Km$) invariant mass resolution
of approximately 6\mevcc.
No veto is found to be necessary to suppress backgrounds from \B decays to charmonia
and a $\pip\pim$ pair final states.

Further separation between signal and combinatorial background candidates relies on
MLPs implemented with the TMVA toolkit~\cite{TMVA}.
The MLPs are trained using simulated \BdPLbarPi samples,
generated according to a constant matrix element without intermediate resonances,
to represent the signal, and with data
from the high-mass sideband region $5400 < m(\PLbarPi) < 5600\mevcc$ for the background,
to avoid partially reconstructed backgrounds.
Separate MLPs are trained and optimized for each year of data taking
and for the two \Lz reconstruction categories. Each MLP is used to select both \BdPLbarPi
and \BsPLbarK candidates.

The seventeen variables used in the MLP classifiers are properties of the $B$ candidate,
the charged hadrons and the \Lz decay products. The input variables are the following:
the \chisq per degree of freedom of the kinematic fit of the decay chain~\cite{Hulsbergen:2005pu};
the IP for all particles calculated with respect to the associated PV;
the distance of closest approach between the two charged hadrons and
the sum of their corresponding \chisqip;
the \Lz candidate decay-length significance with respect to the $B$ vertex, \ie the decay length
divided by its uncertainty;
the angle between the \Lz momentum and the spacial vector connecting the \B and \Lz decay vertices
in the \B rest frame; the \Lz decay time;
the $B$-meson \pt, pseudorapidity, direction angle $\theta_B$, decay-length significance and decay time;
the \Lz helicity angle defined by the \Lz momentum in the \B rest frame and the boost axis
of the \B meson, which is given by the \B-meson momentum in the laboratory frame;
the pointing variable defined as
$
P = [\sum_{p,\Lbar,h^-}{p}\times \sin{\theta_B}]/[\sum_{p,\Lbar,h^-}{p}\times \sin{\theta_B} + \sum_{p,\Lbar,h^-}{\pt}]
$.
The optimal MLP requirement for each of the four subsamples is determined by maximizing the
signal significance of the \BdPLbarPi normalization decay,
with the variation of the signal efficiency with MLP cut value determined from simulation.

A PID selection is applied to the charged hadrons after the MLP selection. No additional PID requirement
is applied to the proton from the \Lz candidate since no contamination from
misidentified \KSPiPi decays is observed.
The optimization of the PID requirements follows the same procedure as the optimization
of the MLP selection.
If more than one candidate is selected in any event of any subsample,
which occurs in about 5\% of selected events, one is chosen at random.

Large data control samples of $\Dz \to \Km \pip$, \LPPi  and \LcPKPi decays
are employed~\cite{LHCb-DP-2012-003} to determine the efficiency of the PID requirements.
All other selection efficiencies are determined from simulation.
It is necessary to account for the distribution of signal candidates
and the variation of the efficiency over the phase space of the decay.
The variation is well described by the
factorized efficiencies in the two-dimensional space of the variables
$m^2(p \Lbar)$ and $m^2(p h^-)$ defining the Dalitz plot.
Simulated events are binned in $m^2(p \Lbar)$ in order to determine the selection efficiencies,
the variation in $m^2(p h^-)$ being mild and therefore integrated out.
The distribution of signal decays in the phase space is obtained
separately for each spectrum with the \sPlot technique~\cite{Pivk:2004ty}
with the $B$-meson candidate invariant mass used as the discriminating variable.
The overall efficiencies of this analysis are of order $10^{-4}$.

The efficiency of the software trigger selection on both decay
modes varied during the data-taking period.
During the 2011 data taking, downstream tracks were not reconstructed in the software
trigger. Such tracks were included in the trigger during the 2012
data taking and a further significant improvement in the algorithms was
implemented mid-year. 
The corresponding changes to the trigger efficiency are taken into account.

Potential sources of background to the \PLbarH spectra are investigated
using simulation samples.
Cross-feed between the \BdPLbarPi and \BsPLbarK decay modes is the dominant source of peaking background.
The loop-mediated decays \BdPLbarK and \BsPLbarPi are suppressed and estimated to be insignificant~\cite{Geng2017205}.
Pion-kaon misidentification from $b$-baryon decays such as the recently observed decays
\LbLHH~\cite{LHCb-PAPER-2016-004} is found to be negligible.
The influence of proton-pion misidentification in the reconstruction and selection
of the \Lz baryon arising from \KS cross-feed is checked
since the PID requirement on the proton from the \Lz is rather loose.
It is verified with Armenteros-Podolanski plots~\cite{Podolanski:1954}
that the \KS contamination can be ignored.
Cross-feed from the presently unobserved decay \LbLPPbar due to proton-pion and proton-kaon misidentification 
is assumed to be negligible considering that the proton misidentification rate is small.
Partially reconstructed decays such as the unobserved
\BdPLbarRho and \BsPLbarKstar modes are treated as a source of systematic uncertainty.
Decay modes containing a \Sigmazbar baryon decaying into $\Sigmazbar\to\Lbar\gamma$,
where the $\gamma$ is not detected, can pollute the signal regions
due to the small mass difference $m(\Sigmaz) - m(\Lz) \approx 77\mevcc$~\cite{PDG2016}.
The decay \BdPSbarPi is expected to have a branching fraction at the level of $10^{-6}$~\cite{Chua:2002wn},
though searches for the \BPSbarH family of decays have found no signal~\cite{Wang:2003yi}.
The decays \BdPSbarPi and \BsPSbarK are expected to be the dominant members of the family and are included
in the fits to the data.

The yields of the signal and background candidates in eight subsamples are determined
from a simultaneous unbinned extended maximum likelihood fit
to the \PLbarH invariant mass distributions.
The eight subsamples correspond to the 2011 and 2012 data-taking periods,
the two \Lz reconstruction categories, and the \PLbarPi and \PLbarK final state hypotheses.
This approach allows the use of common shape parameters,
and the level of cross-feed background can be better constrained by fitting
all subsamples simultaneously.
The probability density function in each subsample is defined as
the sum of components accounting for the signal decay, the cross-feed contribution,
the \BdPSbarPi and \BsPSbarK decays, and combinatorial background.

The signal and normalization modes are modeled
with the sum of two Novosibirsk functions~\cite{Aubert:2002pk}.
All shape parameters are fixed to the values obtained
separately for each subsample from simulation samples.
The \BdPLbarPi and \BsPLbarK peak positions are free parameters determined simultaneously in
all subsamples.
The cross-feed \BsPLbarK (\BdPLbarPi) in the \PLbarPi (\PLbarK) invariant mass distribution
is modeled with the sum of a Gaussian and a modified Fermi function
defined as the product of an exponential and a Fermi-Dirac function. The \BdPSbarPi and \BsPSbarK
decays are modeled differently according to the \Lz reconstruction category and the \PLbarH invariant
mass hypothesis under which they are reconstructed. Depending on the category a modified Fermi function,
a sum of two Novosibirsk functions, the sum of a Novosibirsk and a Gaussian function,
or the sum of a Novosibirsk and a modified Fermi function are used. A combinatorial background component
described by an exponential function is present for both \PLbarH final states.

The yields of the \BdPLbarPi candidates are determined in the fit together with the ratio
of the \BsPLbarK to \BdPLbarPi branching fractions, which is determined simultaneously
across all subsamples accounting for differences in selection efficiencies. These depend
on the data-taking period, \Lz reconstruction category and mass hypothesis of the meson from the $B$ decay.
The uncertainties arising from the ratios of efficiencies are included in the fit as Gaussian constraints.
The yields of the \BdPSbarPi and \BsPSbarK decays are defined relative to those of the corresponding
\BdPLbarPi and \BsPLbarK decays, respectively.
These two $\Sigmaz$-to-$\Lz$ decay yield ratios are determined simultaneously in the fit
across all subsamples following the same procedure as for the \BsPLbarK decay.
The combinatorial background yield and shape parameters are treated independently in each subsample
and are allowed to vary in the fit.

\begin{figure}[tb]
  \centering
  \includegraphics[width=0.495\textwidth]{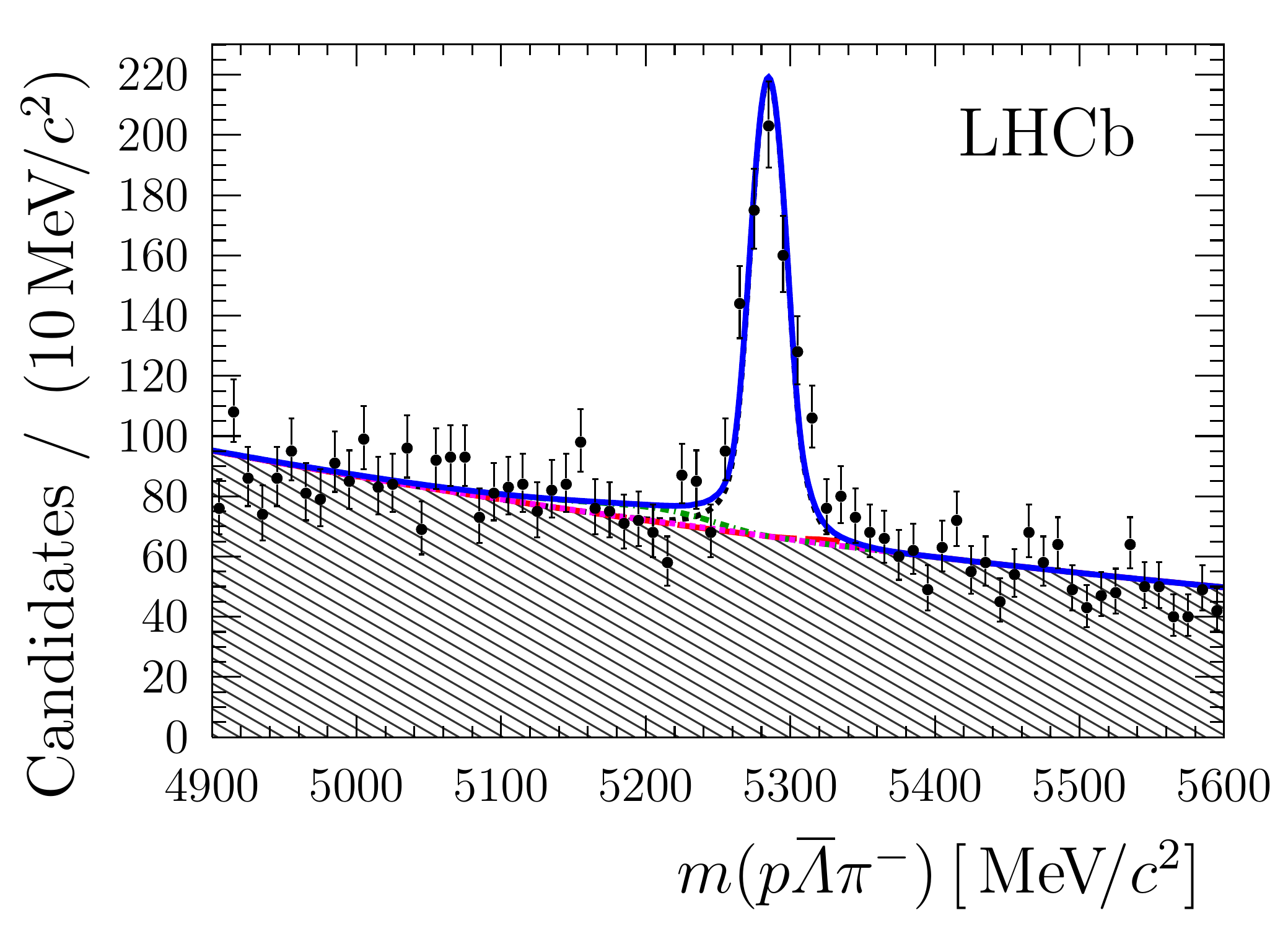}
  \includegraphics[width=0.495\textwidth]{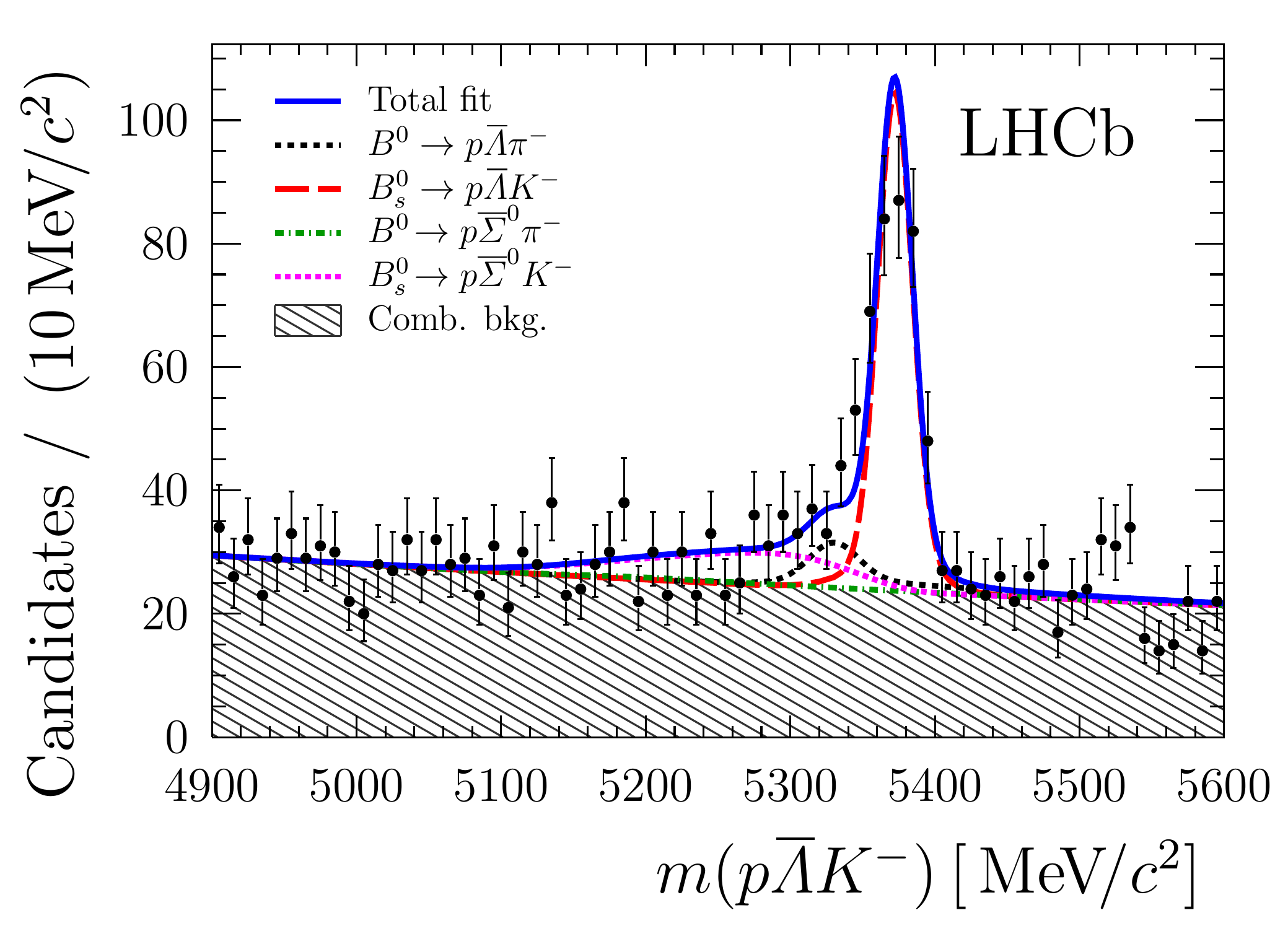}
  \caption{Mass distributions for $b$-hadron candidates for
           (left) the \PLbarPi and (right) the \PLbarK sample for the
           combined long and downstream categories. The black points represent the data,
           the solid blue curve the result of the fit, the red dashed curve
           the \BsPLbarK contribution, the black (magenta) dotted curve the \BdPLbarPi (\BsPSbarK)
           and the green dash-dotted curve the contribution from \BdPSbarPi decays. The combinatorial
           background distribution is indicated by the shaded histogram.}
  \label{fig:DataFit}
\end{figure}

Figure~\ref{fig:DataFit} presents the fit
to the \PLbarH invariant mass distributions for all subsamples combined.
Both \BdPLbarPi and \BsPLbarK signals are prominent.
In particular, the \BsPLbarK decay is observed with a statistical significance
above 15 standard deviations, estimated from the change in log-likelihood between fits
with and without the \BsPLbarK signal component~\cite{Wilks:1938dza}.
It constitutes the first observation of a baryonic \Bs decay.
The yields summed over all subsamples are $N(\BdPLbarPi) = 519 \pm 28$
and $N(\BsPLbarK) = 234 \pm 29$, where the uncertainties are statistical only.

\begin{figure}[tb]
  \centering
  \includegraphics[width=0.495\textwidth]{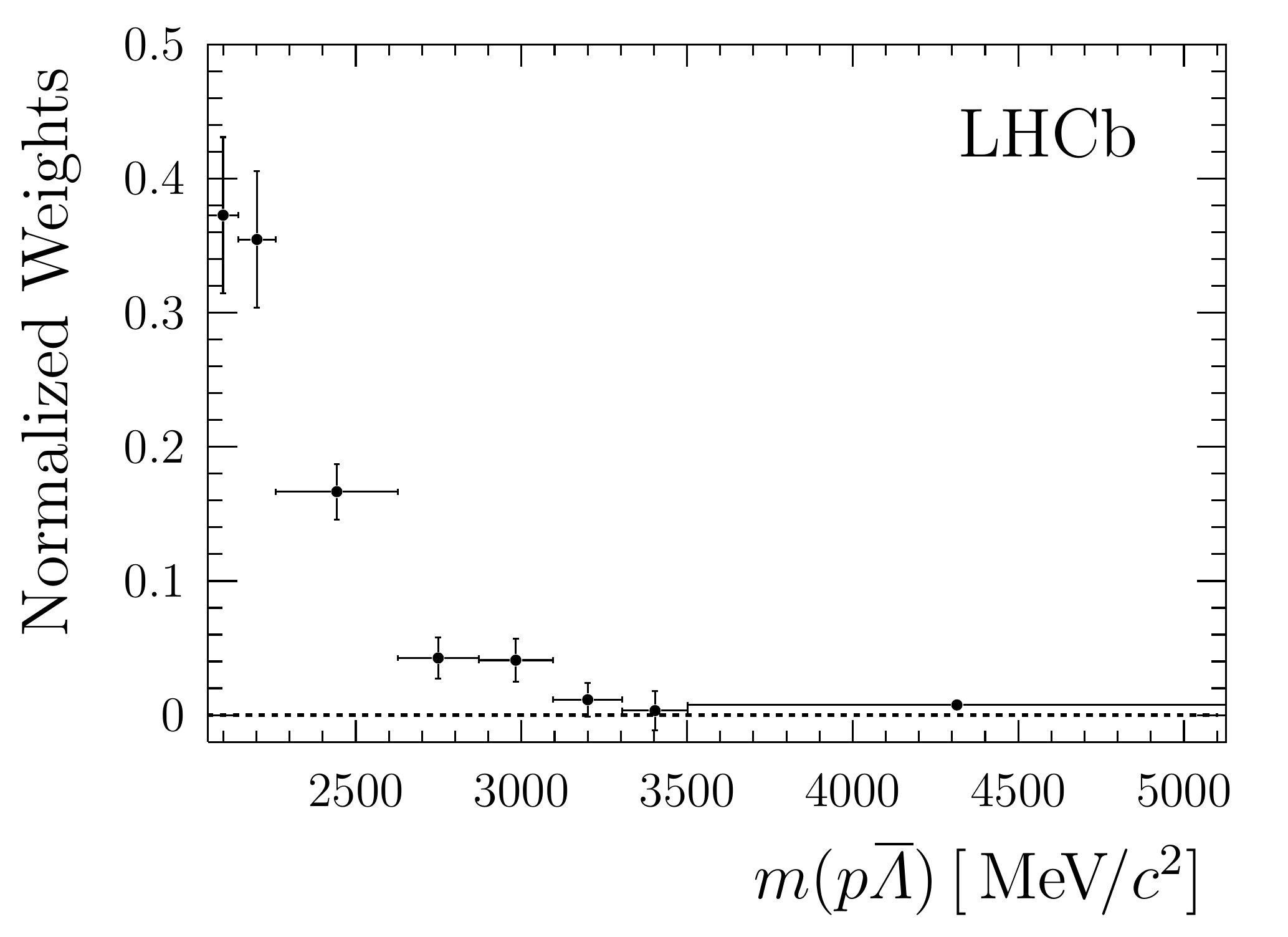}
  \includegraphics[width=0.495\textwidth]{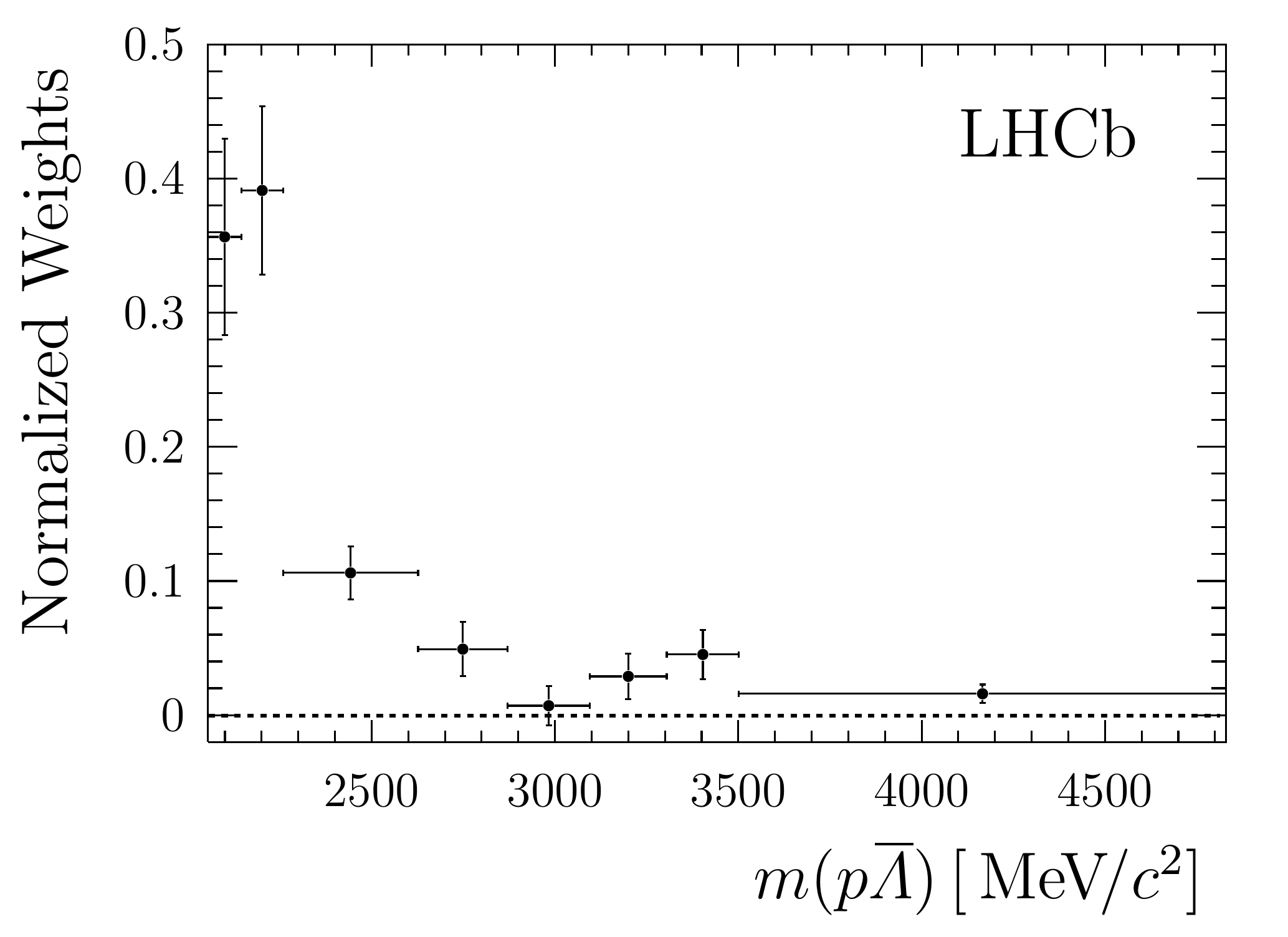}
  \caption{Efficiency-corrected and background-subtracted $m(\PLbar)$ invariant mass
           distributions for (left) \BdPLbarPi and (right) \BsPLbarK candidates.
           The distributions are normalized to unity.}
  \label{fig:2bodyMasses}
\end{figure}

The \sPlot technique is used to subtract the background and obtain the phase space distribution of signal candidates.
Figure~\ref{fig:2bodyMasses} shows the $m(\PLbar)$ invariant mass distributions
for the \BdPLbarPi and \BsPLbarK candidates after correcting for the distribution selection efficiencies.
Both distributions show a pronounced enhancement at threshold in the
baryon-antibaryon invariant mass, first suggested in Ref.~\cite{Hou:2000bz} and observed in several baryonic \B decay modes.

The sources of systematic uncertainty
arise from the fit model, the knowledge of the selection efficiencies,
and the uncertainties on the \BdPLbarPi branching fraction
and on the ratio of hadronization probabilities $f_s/f_d$.
Uncertainties on the selection efficiencies arise from
residual differences between data and simulation in the trigger, reconstruction,
selection and particle identification. Additional uncertainties arise due to the limited size of the
simulation samples and the corresponding uncertainty on the distribution of the efficiencies across the decay phase space.
As the efficiencies depend on the signal decay-time distribution, the effect coming
from the different lifetimes of the \Bs mass eigenstates has been evaluated~\cite{DeBruyn:2012wj}.
Pseudoexperiments are used to estimate the effect of using alternative shapes
for the fit components, of including additional backgrounds in the fit such as
partially reconstructed decays,
and of excluding the \BdPSbarPi and \BsPSbarK decays that show no significant contribution.
Intrinsic biases in the fitted signal yields are investigated with ensembles of
simulated pseudoexperiments. A small bias is found and added to the systematic uncertainty on the fit model.
The systematic uncertainty due to the knowledge of the efficiencies
involved in the definition of fit constraints is negligible.
The total systematic uncertainty on the \BsPLbarK branching fraction is given by the sum
of all uncertainties added in quadrature and amounts to 10.5\%;
it is dominated by the systematic uncertainty on the fit model.

The uncertainty on the branching fraction of the normalization decay,
\mbox{$\BF(\BdPLbarPi) = (3.14 \pm 0.29) \times 10^{-6}$~\cite{PDG2016}},
is taken as a systematic uncertainty from external inputs.
The 5.8\% uncertainty on the latest $f_s/f_d$ combination from \lhcb,
$f_s/f_d = 0.259 \pm 0.015$~\cite{fsfd},
is taken as a second source of systematic uncertainty from external inputs.

The \BsPLbarK branching fraction, determined relative to that of the
\BdPLbarPi normalization channel according to Eq.~\ref{eq:BF},
is measured to be
$$
\BF(\BsPLbarK) + \BF(\BsPbarLK) = \left[\vphantom{\sum}{5.46} \pm{0.61} \pm 0.57 \pm 0.50 (\BR) \pm 0.32 (f_s/f_d)\right]\times 10^{-6} \;,
$$
where the first uncertainty is statistical and the second systematic,
the third uncertainty accounts for the experimental uncertainty on the
branching fraction of the \BdPLbarPi decay,
and the fourth uncertainty relates to the knowledge of $f_s/f_d$.

In summary, the first observation of the three-body charmless baryonic decay
\mbox{\BsPLbarK} is reported using a proton-proton collision data sample collected by the \lhcb experiment,
corresponding to an integrated luminosity of 3.0\invfb.
The decay is observed with a statistical significance above 15 standard deviations,
which constitutes the first observation of a baryonic \Bs decay.

Decays of $B$ mesons to final states containing baryons are now observed
for all $B$-meson species. Their study provides valuable information on the
dynamics of hadronic decays of $B$ mesons.
The present analysis motivates further theoretical studies of baryonic \Bs decays
in addition to those currently
published~\cite{Geng2017205,Hsiao:2014tda,PhysRevD.89.056003,Hsiao:2014zza}.

\section*{Acknowledgements}

\noindent We express our gratitude to our colleagues in the CERN
accelerator departments for the excellent performance of the LHC. We
thank the technical and administrative staff at the LHCb
institutes. We acknowledge support from CERN and from the national
agencies: CAPES, CNPq, FAPERJ and FINEP (Brazil); MOST and NSFC (China);
CNRS/IN2P3 (France); BMBF, DFG and MPG (Germany); INFN (Italy); 
NWO (The Netherlands); MNiSW and NCN (Poland); MEN/IFA (Romania); 
MinES and FASO (Russia); MinECo (Spain); SNSF and SER (Switzerland); 
NASU (Ukraine); STFC (United Kingdom); NSF (USA).
We acknowledge the computing resources that are provided by CERN, IN2P3 (France), KIT and DESY (Germany), INFN (Italy), SURF (The Netherlands), PIC (Spain), GridPP (United Kingdom), RRCKI and Yandex LLC (Russia), CSCS (Switzerland), IFIN-HH (Romania), CBPF (Brazil), PL-GRID (Poland) and OSC (USA). We are indebted to the communities behind the multiple open 
source software packages on which we depend.
Individual groups or members have received support from AvH Foundation (Germany),
EPLANET, Marie Sk\l{}odowska-Curie Actions and ERC (European Union), 
Conseil G\'{e}n\'{e}ral de Haute-Savoie, Labex ENIGMASS and OCEVU, 
R\'{e}gion Auvergne (France), RFBR and Yandex LLC (Russia), GVA, XuntaGal and GENCAT (Spain), Herchel Smith Fund, The Royal Society, Royal Commission for the Exhibition of 1851 and the Leverhulme Trust (United Kingdom).

\addcontentsline{toc}{section}{References}
\setboolean{inbibliography}{true}
\bibliographystyle{LHCb}
\bibliography{main,LHCb-PAPER,LHCb-CONF,LHCb-DP,LHCb-TDR}

\ifx\mcitethebibliography\mciteundefinedmacro
\PackageError{LHCb.bst}{mciteplus.sty has not been loaded}
{This bibstyle requires the use of the mciteplus package.}\fi
\providecommand{\href}[2]{#2}
\begin{mcitethebibliography}{10}
\mciteSetBstSublistMode{n}
\mciteSetBstMaxWidthForm{subitem}{\alph{mcitesubitemcount})}
\mciteSetBstSublistLabelBeginEnd{\mcitemaxwidthsubitemform\space}
{\relax}{\relax}

\bibitem{PhysRevLett.79.3125}
\cleo collaboration, X.~Fu,  {\em et~al.},
  \ifthenelse{\boolean{articletitles}}{\emph{{Observation of exclusive $B$
  decays to final states containing a charmed baryon}},
  }{}\href{http://dx.doi.org/10.1103/PhysRevLett.79.3125}{Phys.\ Rev.\ Lett.\
  \textbf{79} (1997) 3125}\relax
\mciteBstWouldAddEndPuncttrue
\mciteSetBstMidEndSepPunct{\mcitedefaultmidpunct}
{\mcitedefaultendpunct}{\mcitedefaultseppunct}\relax
\EndOfBibitem
\bibitem{Bevan:2014iga}
\babar and \belle collaborations, A.~J. Bevan {\em et~al.},
  \ifthenelse{\boolean{articletitles}}{\emph{{The physics of the B factories}},
  }{}\href{http://dx.doi.org/10.1140/epjc/s10052-014-3026-9}{Eur.\ Phys.\ J.\
  \textbf{C74} (2014) 3026},
  \href{http://arxiv.org/abs/1406.6311}{{\normalfont\ttfamily
  arXiv:1406.6311}}\relax
\mciteBstWouldAddEndPuncttrue
\mciteSetBstMidEndSepPunct{\mcitedefaultmidpunct}
{\mcitedefaultendpunct}{\mcitedefaultseppunct}\relax
\EndOfBibitem
\bibitem{LHCb-PAPER-2014-039}
LHCb collaboration, R.~Aaij {\em et~al.},
  \ifthenelse{\boolean{articletitles}}{\emph{{First observation of a baryonic
  $\Bcp$ decay}},
  }{}\href{http://dx.doi.org/10.1103/PhysRevLett.113.152003}{Phys.\ Rev.\
  Lett.\  \textbf{113} (2014) 152003},
  \href{http://arxiv.org/abs/1408.0971}{{\normalfont\ttfamily
  arXiv:1408.0971}}\relax
\mciteBstWouldAddEndPuncttrue
\mciteSetBstMidEndSepPunct{\mcitedefaultmidpunct}
{\mcitedefaultendpunct}{\mcitedefaultseppunct}\relax
\EndOfBibitem
\bibitem{Solovieva:2013rhq}
\belle collaboration, E.~Solovieva {\em et~al.},
  \ifthenelse{\boolean{articletitles}}{\emph{{Evidence for $\Bsb \to \Lc \Lbar
  \pim$}}, }{}\href{http://dx.doi.org/10.1016/j.physletb.2013.08.057}{Phys.\
  Lett.\  \textbf{B 726} (2013) 206},
  \href{http://arxiv.org/abs/1304.6931}{{\normalfont\ttfamily
  arXiv:1304.6931}}\relax
\mciteBstWouldAddEndPuncttrue
\mciteSetBstMidEndSepPunct{\mcitedefaultmidpunct}
{\mcitedefaultendpunct}{\mcitedefaultseppunct}\relax
\EndOfBibitem
\bibitem{Hou:2000bz}
W.-S. Hou and A.~Soni, \ifthenelse{\boolean{articletitles}}{\emph{{Pathways to
  rare baryonic $B$ decays}},
  }{}\href{http://dx.doi.org/10.1103/PhysRevLett.86.4247}{Phys.\ Rev.\ Lett.\
  \textbf{86} (2001) 4247},
  \href{http://arxiv.org/abs/hep-ph/0008079}{{\normalfont\ttfamily
  arXiv:hep-ph/0008079}}\relax
\mciteBstWouldAddEndPuncttrue
\mciteSetBstMidEndSepPunct{\mcitedefaultmidpunct}
{\mcitedefaultendpunct}{\mcitedefaultseppunct}\relax
\EndOfBibitem
\bibitem{Hsiao:2014zza}
Y.~K. Hsiao and C.~Q. Geng,
  \ifthenelse{\boolean{articletitles}}{\emph{{Violation of partial conservation
  of the axial-vector current and two-body baryonic $B$ and $D_s$ decays}},
  }{}\href{http://dx.doi.org/10.1103/PhysRevD.91.077501}{Phys.\ Rev.\
  \textbf{D91} (2015) 077501},
  \href{http://arxiv.org/abs/1407.7639}{{\normalfont\ttfamily
  arXiv:1407.7639}}\relax
\mciteBstWouldAddEndPuncttrue
\mciteSetBstMidEndSepPunct{\mcitedefaultmidpunct}
{\mcitedefaultendpunct}{\mcitedefaultseppunct}\relax
\EndOfBibitem
\bibitem{Cheng:2014qxa}
H.-Y. Cheng and C.-K. Chua, \ifthenelse{\boolean{articletitles}}{\emph{{On the
  smallness of tree-dominated charmless two-body baryonic $B$ decay rates}},
  }{}\href{http://dx.doi.org/10.1103/PhysRevD.91.036003}{Phys.\ Rev.\
  \textbf{D91} (2015) 036003},
  \href{http://arxiv.org/abs/1412.8272}{{\normalfont\ttfamily
  arXiv:1412.8272}}\relax
\mciteBstWouldAddEndPuncttrue
\mciteSetBstMidEndSepPunct{\mcitedefaultmidpunct}
{\mcitedefaultendpunct}{\mcitedefaultseppunct}\relax
\EndOfBibitem
\bibitem{Geng2017205}
C.~Q. Geng, Y.~K. Hsiao, and E.~Rodrigues,
  \ifthenelse{\boolean{articletitles}}{\emph{{Three-body charmless baryonic
  \Bsb decays}},
  }{}\href{http://dx.doi.org/10.1016/j.physletb.2017.02.001}{Phys.\ Lett.\
  \textbf{B 767} (2017) 205},
  \href{http://arxiv.org/abs/1612.08133}{{\normalfont\ttfamily
  arXiv:1612.08133}}\relax
\mciteBstWouldAddEndPuncttrue
\mciteSetBstMidEndSepPunct{\mcitedefaultmidpunct}
{\mcitedefaultendpunct}{\mcitedefaultseppunct}\relax
\EndOfBibitem
\bibitem{Aubert:2009am}
\babar collaboration, B.~Aubert {\em et~al.},
  \ifthenelse{\boolean{articletitles}}{\emph{{Measurement of the branching
  fraction and \Lbar polarization in $\Bz \to \Lbar \proton \pim$}},
  }{}\href{http://dx.doi.org/10.1103/PhysRevD.79.112009}{Phys.\ Rev.\
  \textbf{D79} (2009) 112009},
  \href{http://arxiv.org/abs/0904.4724}{{\normalfont\ttfamily
  arXiv:0904.4724}}\relax
\mciteBstWouldAddEndPuncttrue
\mciteSetBstMidEndSepPunct{\mcitedefaultmidpunct}
{\mcitedefaultendpunct}{\mcitedefaultseppunct}\relax
\EndOfBibitem
\bibitem{Wang:2003yi}
\belle collaboration, M.~Z. Wang {\em et~al.},
  \ifthenelse{\boolean{articletitles}}{\emph{{Observation of $\Bd \to \proton
  \Lbar \pim$}},
  }{}\href{http://dx.doi.org/10.1103/PhysRevLett.90.201802}{Phys.\ Rev.\ Lett.\
   \textbf{90} (2003) 201802},
  \href{http://arxiv.org/abs/hep-ex/0302024}{{\normalfont\ttfamily
  arXiv:hep-ex/0302024}}\relax
\mciteBstWouldAddEndPuncttrue
\mciteSetBstMidEndSepPunct{\mcitedefaultmidpunct}
{\mcitedefaultendpunct}{\mcitedefaultseppunct}\relax
\EndOfBibitem
\bibitem{Wang:2007as}
\belle collaboration, M.-Z. Wang {\em et~al.},
  \ifthenelse{\boolean{articletitles}}{\emph{{Study of $\Bp \to p \Lbar \gamma,
  p \Lbar \piz$ and $\Bz \to p \Lbar \pim$}},
  }{}\href{http://dx.doi.org/10.1103/PhysRevD.76.052004}{Phys.\ Rev.\
  \textbf{D76} (2007) 052004},
  \href{http://arxiv.org/abs/0704.2672}{{\normalfont\ttfamily
  arXiv:0704.2672}}\relax
\mciteBstWouldAddEndPuncttrue
\mciteSetBstMidEndSepPunct{\mcitedefaultmidpunct}
{\mcitedefaultendpunct}{\mcitedefaultseppunct}\relax
\EndOfBibitem
\bibitem{Geng:2008ps}
C.~Q. Geng and Y.~K. Hsiao, \ifthenelse{\boolean{articletitles}}{\emph{{Direct
  \CP and $T$ violation in baryonic $B$ decays}},
  }{}\href{http://dx.doi.org/10.1142/S0217751X08041992}{Int.\ J.\ Mod.\ Phys.\
  \textbf{A23} (2008) 3290},
  \href{http://arxiv.org/abs/0801.0022}{{\normalfont\ttfamily
  arXiv:0801.0022}}\relax
\mciteBstWouldAddEndPuncttrue
\mciteSetBstMidEndSepPunct{\mcitedefaultmidpunct}
{\mcitedefaultendpunct}{\mcitedefaultseppunct}\relax
\EndOfBibitem
\bibitem{Alves:2008zz}
LHCb collaboration, A.~A. Alves~Jr.\ {\em et~al.},
  \ifthenelse{\boolean{articletitles}}{\emph{{The \lhcb detector at the LHC}},
  }{}\href{http://dx.doi.org/10.1088/1748-0221/3/08/S08005}{JINST \textbf{3}
  (2008) S08005}\relax
\mciteBstWouldAddEndPuncttrue
\mciteSetBstMidEndSepPunct{\mcitedefaultmidpunct}
{\mcitedefaultendpunct}{\mcitedefaultseppunct}\relax
\EndOfBibitem
\bibitem{LHCb-DP-2014-002}
LHCb collaboration, R.~Aaij {\em et~al.},
  \ifthenelse{\boolean{articletitles}}{\emph{{LHCb detector performance}},
  }{}\href{http://dx.doi.org/10.1142/S0217751X15300227}{Int.\ J.\ Mod.\ Phys.\
  \textbf{A30} (2015) 1530022},
  \href{http://arxiv.org/abs/1412.6352}{{\normalfont\ttfamily
  arXiv:1412.6352}}\relax
\mciteBstWouldAddEndPuncttrue
\mciteSetBstMidEndSepPunct{\mcitedefaultmidpunct}
{\mcitedefaultendpunct}{\mcitedefaultseppunct}\relax
\EndOfBibitem
\bibitem{Sjostrand:2007gs}
T.~Sj\"{o}strand, S.~Mrenna, and P.~Skands,
  \ifthenelse{\boolean{articletitles}}{\emph{{A brief introduction to PYTHIA
  8.1}}, }{}\href{http://dx.doi.org/10.1016/j.cpc.2008.01.036}{Comput.\ Phys.\
  Commun.\  \textbf{178} (2008) 852},
  \href{http://arxiv.org/abs/0710.3820}{{\normalfont\ttfamily
  arXiv:0710.3820}}\relax
\mciteBstWouldAddEndPuncttrue
\mciteSetBstMidEndSepPunct{\mcitedefaultmidpunct}
{\mcitedefaultendpunct}{\mcitedefaultseppunct}\relax
\EndOfBibitem
\bibitem{Sjostrand:2006za}
T.~Sj\"{o}strand, S.~Mrenna, and P.~Skands,
  \ifthenelse{\boolean{articletitles}}{\emph{{PYTHIA 6.4 physics and manual}},
  }{}\href{http://dx.doi.org/10.1088/1126-6708/2006/05/026}{JHEP \textbf{05}
  (2006) 026}, \href{http://arxiv.org/abs/hep-ph/0603175}{{\normalfont\ttfamily
  arXiv:hep-ph/0603175}}\relax
\mciteBstWouldAddEndPuncttrue
\mciteSetBstMidEndSepPunct{\mcitedefaultmidpunct}
{\mcitedefaultendpunct}{\mcitedefaultseppunct}\relax
\EndOfBibitem
\bibitem{LHCb-PROC-2010-056}
I.~Belyaev {\em et~al.}, \ifthenelse{\boolean{articletitles}}{\emph{{Handling
  of the generation of primary events in Gauss, the LHCb simulation
  framework}}, }{}\href{http://dx.doi.org/10.1088/1742-6596/331/3/032047}{{J.\
  Phys.\ Conf.\ Ser.\ } \textbf{331} (2011) 032047}\relax
\mciteBstWouldAddEndPuncttrue
\mciteSetBstMidEndSepPunct{\mcitedefaultmidpunct}
{\mcitedefaultendpunct}{\mcitedefaultseppunct}\relax
\EndOfBibitem
\bibitem{Lange:2001uf}
D.~J. Lange, \ifthenelse{\boolean{articletitles}}{\emph{{The EvtGen particle
  decay simulation package}},
  }{}\href{http://dx.doi.org/10.1016/S0168-9002(01)00089-4}{Nucl.\ Instrum.\
  Meth.\  \textbf{A462} (2001) 152}\relax
\mciteBstWouldAddEndPuncttrue
\mciteSetBstMidEndSepPunct{\mcitedefaultmidpunct}
{\mcitedefaultendpunct}{\mcitedefaultseppunct}\relax
\EndOfBibitem
\bibitem{Golonka:2005pn}
P.~Golonka and Z.~Was, \ifthenelse{\boolean{articletitles}}{\emph{{PHOTOS Monte
  Carlo: A precision tool for QED corrections in $Z$ and $W$ decays}},
  }{}\href{http://dx.doi.org/10.1140/epjc/s2005-02396-4}{Eur.\ Phys.\ J.\
  \textbf{C45} (2006) 97},
  \href{http://arxiv.org/abs/hep-ph/0506026}{{\normalfont\ttfamily
  arXiv:hep-ph/0506026}}\relax
\mciteBstWouldAddEndPuncttrue
\mciteSetBstMidEndSepPunct{\mcitedefaultmidpunct}
{\mcitedefaultendpunct}{\mcitedefaultseppunct}\relax
\EndOfBibitem
\bibitem{Allison:2006ve}
Geant4 collaboration, J.~Allison {\em et~al.},
  \ifthenelse{\boolean{articletitles}}{\emph{{Geant4 developments and
  applications}}, }{}\href{http://dx.doi.org/10.1109/TNS.2006.869826}{IEEE
  Trans.\ Nucl.\ Sci.\  \textbf{53} (2006) 270}\relax
\mciteBstWouldAddEndPuncttrue
\mciteSetBstMidEndSepPunct{\mcitedefaultmidpunct}
{\mcitedefaultendpunct}{\mcitedefaultseppunct}\relax
\EndOfBibitem
\bibitem{Agostinelli:2002hh}
Geant4 collaboration, S.~Agostinelli {\em et~al.},
  \ifthenelse{\boolean{articletitles}}{\emph{{Geant4: A simulation toolkit}},
  }{}\href{http://dx.doi.org/10.1016/S0168-9002(03)01368-8}{Nucl.\ Instrum.\
  Meth.\  \textbf{A506} (2003) 250}\relax
\mciteBstWouldAddEndPuncttrue
\mciteSetBstMidEndSepPunct{\mcitedefaultmidpunct}
{\mcitedefaultendpunct}{\mcitedefaultseppunct}\relax
\EndOfBibitem
\bibitem{LHCb-PROC-2011-006}
M.~Clemencic {\em et~al.}, \ifthenelse{\boolean{articletitles}}{\emph{{The
  \lhcb simulation application, Gauss: Design, evolution and experience}},
  }{}\href{http://dx.doi.org/10.1088/1742-6596/331/3/032023}{{J.\ Phys.\ Conf.\
  Ser.\ } \textbf{331} (2011) 032023}\relax
\mciteBstWouldAddEndPuncttrue
\mciteSetBstMidEndSepPunct{\mcitedefaultmidpunct}
{\mcitedefaultendpunct}{\mcitedefaultseppunct}\relax
\EndOfBibitem
\bibitem{LHCb-DP-2012-004}
R.~Aaij {\em et~al.}, \ifthenelse{\boolean{articletitles}}{\emph{{The \lhcb
  trigger and its performance in 2011}},
  }{}\href{http://dx.doi.org/10.1088/1748-0221/8/04/P04022}{JINST \textbf{8}
  (2013) P04022}, \href{http://arxiv.org/abs/1211.3055}{{\normalfont\ttfamily
  arXiv:1211.3055}}\relax
\mciteBstWouldAddEndPuncttrue
\mciteSetBstMidEndSepPunct{\mcitedefaultmidpunct}
{\mcitedefaultendpunct}{\mcitedefaultseppunct}\relax
\EndOfBibitem
\bibitem{BBDT}
V.~V. Gligorov and M.~Williams,
  \ifthenelse{\boolean{articletitles}}{\emph{{Efficient, reliable and fast
  high-level triggering using a bonsai boosted decision tree}},
  }{}\href{http://dx.doi.org/10.1088/1748-0221/8/02/P02013}{JINST \textbf{8}
  (2013) P02013}, \href{http://arxiv.org/abs/1210.6861}{{\normalfont\ttfamily
  arXiv:1210.6861}}\relax
\mciteBstWouldAddEndPuncttrue
\mciteSetBstMidEndSepPunct{\mcitedefaultmidpunct}
{\mcitedefaultendpunct}{\mcitedefaultseppunct}\relax
\EndOfBibitem
\bibitem{MLPs}
D.~E. Rumelhart, G.~E. Hinton, and R.~J. Williams, {\em Parallel distributed
  processing: explorations in the microstructure of cognition}, vol.~1, MIT,
  Cambridge, USA, 1986\relax
\mciteBstWouldAddEndPuncttrue
\mciteSetBstMidEndSepPunct{\mcitedefaultmidpunct}
{\mcitedefaultendpunct}{\mcitedefaultseppunct}\relax
\EndOfBibitem
\bibitem{Hulsbergen:2005pu}
W.~D. Hulsbergen, \ifthenelse{\boolean{articletitles}}{\emph{{Decay chain
  fitting with a Kalman filter}},
  }{}\href{http://dx.doi.org/10.1016/j.nima.2005.06.078}{Nucl.\ Instrum.\
  Meth.\  \textbf{A552} (2005) 566},
  \href{http://arxiv.org/abs/physics/0503191}{{\normalfont\ttfamily
  arXiv:physics/0503191}}\relax
\mciteBstWouldAddEndPuncttrue
\mciteSetBstMidEndSepPunct{\mcitedefaultmidpunct}
{\mcitedefaultendpunct}{\mcitedefaultseppunct}\relax
\EndOfBibitem
\bibitem{PDG2016}
Particle Data Group, C.~Patrignani {\em et~al.},
  \ifthenelse{\boolean{articletitles}}{\emph{{\href{http://pdg.lbl.gov/}{Review
  of particle physics}}},
  }{}\href{http://dx.doi.org/10.1088/1674-1137/40/10/100001}{Chin.\ Phys.\
  \textbf{C40} (2016) 100001}\relax
\mciteBstWouldAddEndPuncttrue
\mciteSetBstMidEndSepPunct{\mcitedefaultmidpunct}
{\mcitedefaultendpunct}{\mcitedefaultseppunct}\relax
\EndOfBibitem
\bibitem{TMVA}
P.~Speckmayer, A.~Hoecker, J.~Stelzer, and H.~Voss,
  \ifthenelse{\boolean{articletitles}}{\emph{{The toolkit for multivariate data
  analysis: TMVA 4}},
  }{}\href{http://dx.doi.org/10.1088/1742-6596/219/3/032057}{J.\ Phys.\ Conf.\
  Ser.\  \textbf{219} (2010) 032057}\relax
\mciteBstWouldAddEndPuncttrue
\mciteSetBstMidEndSepPunct{\mcitedefaultmidpunct}
{\mcitedefaultendpunct}{\mcitedefaultseppunct}\relax
\EndOfBibitem
\bibitem{LHCb-DP-2012-003}
M.~Adinolfi {\em et~al.},
  \ifthenelse{\boolean{articletitles}}{\emph{{Performance of the \lhcb RICH
  detector at the LHC}},
  }{}\href{http://dx.doi.org/10.1140/epjc/s10052-013-2431-9}{Eur.\ Phys.\ J.\
  \textbf{C73} (2013) 2431},
  \href{http://arxiv.org/abs/1211.6759}{{\normalfont\ttfamily
  arXiv:1211.6759}}\relax
\mciteBstWouldAddEndPuncttrue
\mciteSetBstMidEndSepPunct{\mcitedefaultmidpunct}
{\mcitedefaultendpunct}{\mcitedefaultseppunct}\relax
\EndOfBibitem
\bibitem{Pivk:2004ty}
M.~Pivk and F.~R. Le~Diberder,
  \ifthenelse{\boolean{articletitles}}{\emph{{sPlot: A statistical tool to
  unfold data distributions}},
  }{}\href{http://dx.doi.org/10.1016/j.nima.2005.08.106}{Nucl.\ Instrum.\
  Meth.\  \textbf{A555} (2005) 356},
  \href{http://arxiv.org/abs/physics/0402083}{{\normalfont\ttfamily
  arXiv:physics/0402083}}\relax
\mciteBstWouldAddEndPuncttrue
\mciteSetBstMidEndSepPunct{\mcitedefaultmidpunct}
{\mcitedefaultendpunct}{\mcitedefaultseppunct}\relax
\EndOfBibitem
\bibitem{LHCb-PAPER-2016-004}
LHCb collaboration, R.~Aaij {\em et~al.},
  \ifthenelse{\boolean{articletitles}}{\emph{{Observations of
  $\Lb\to\Lz\Kp\pim$ and $\Lb\to\Lz\Kp\Km$ decays and searches for other $\Lb$
  and $\Xibz$ decays to $\Lz h^+h^-$ final states}},
  }{}\href{http://dx.doi.org/10.1007/JHEP05(2016)081}{JHEP \textbf{05} (2016)
  081}, \href{http://arxiv.org/abs/1603.00413}{{\normalfont\ttfamily
  arXiv:1603.00413}}\relax
\mciteBstWouldAddEndPuncttrue
\mciteSetBstMidEndSepPunct{\mcitedefaultmidpunct}
{\mcitedefaultendpunct}{\mcitedefaultseppunct}\relax
\EndOfBibitem
\bibitem{Podolanski:1954}
J.~Podolanski and R.~Armenteros,
  \ifthenelse{\boolean{articletitles}}{\emph{{III. Analysis of V-events}},
  }{}\href{http://dx.doi.org/10.1080/14786440108520416}{The London, Edinburgh,
  and Dublin Philosophical Magazine and Journal of Science \textbf{45} (1954)
  13}\relax
\mciteBstWouldAddEndPuncttrue
\mciteSetBstMidEndSepPunct{\mcitedefaultmidpunct}
{\mcitedefaultendpunct}{\mcitedefaultseppunct}\relax
\EndOfBibitem
\bibitem{Chua:2002wn}
C.-K. Chua, W.-S. Hou, and S.-Y. Tsai,
  \ifthenelse{\boolean{articletitles}}{\emph{{Charmless three-body baryonic B
  decays}}, }{}\href{http://dx.doi.org/10.1103/PhysRevD.66.054004}{Phys.\ Rev.\
   \textbf{D66} (2002) 054004},
  \href{http://arxiv.org/abs/hep-ph/0204185}{{\normalfont\ttfamily
  arXiv:hep-ph/0204185}}\relax
\mciteBstWouldAddEndPuncttrue
\mciteSetBstMidEndSepPunct{\mcitedefaultmidpunct}
{\mcitedefaultendpunct}{\mcitedefaultseppunct}\relax
\EndOfBibitem
\bibitem{Aubert:2002pk}
\babar collaboration, B.~Aubert {\em et~al.},
  \ifthenelse{\boolean{articletitles}}{\emph{{Search for decays of \Bd mesons
  into pairs of leptons}}, }{} in {\em {Proceedings of the $31^{st}$
  international conference on high energy physics, ICHEP 2002, Amsterdam, The
  Netherlands, July 25-31, 2002}}, 2002.
\newblock \href{http://arxiv.org/abs/hep-ex/0207083}{{\normalfont\ttfamily
  arXiv:hep-ex/0207083}}\relax
\mciteBstWouldAddEndPuncttrue
\mciteSetBstMidEndSepPunct{\mcitedefaultmidpunct}
{\mcitedefaultendpunct}{\mcitedefaultseppunct}\relax
\EndOfBibitem
\bibitem{Wilks:1938dza}
S.~S. Wilks, \ifthenelse{\boolean{articletitles}}{\emph{{The large-sample
  distribution of the likelihood ratio for testing composite hypotheses}},
  }{}\href{http://dx.doi.org/10.1214/aoms/1177732360}{Ann.\ Math.\ Stat.\
  \textbf{9} (1938) 60}\relax
\mciteBstWouldAddEndPuncttrue
\mciteSetBstMidEndSepPunct{\mcitedefaultmidpunct}
{\mcitedefaultendpunct}{\mcitedefaultseppunct}\relax
\EndOfBibitem
\bibitem{DeBruyn:2012wj}
K.~De~Bruyn {\em et~al.}, \ifthenelse{\boolean{articletitles}}{\emph{{Branching
  ratio measurements of \Bs decays}},
  }{}\href{http://dx.doi.org/10.1103/PhysRevD.86.014027}{Phys.\ Rev.\
  \textbf{D86} (2012) 014027},
  \href{http://arxiv.org/abs/1204.1735}{{\normalfont\ttfamily
  arXiv:1204.1735}}\relax
\mciteBstWouldAddEndPuncttrue
\mciteSetBstMidEndSepPunct{\mcitedefaultmidpunct}
{\mcitedefaultendpunct}{\mcitedefaultseppunct}\relax
\EndOfBibitem
\bibitem{fsfd}
LHCb collaboration, R.~Aaij {\em et~al.},
  \ifthenelse{\boolean{articletitles}}{\emph{{Measurement of the fragmentation
  fraction ratio $f_s/f_d$ and its dependence on $B$ meson kinematics}},
  }{}\href{http://dx.doi.org/10.1007/JHEP04(2013)001}{JHEP \textbf{04} (2013)
  001}, \href{http://arxiv.org/abs/1301.5286}{{\normalfont\ttfamily
  arXiv:1301.5286}}, $f_s/f_d$ value updated in
  \href{https://cds.cern.ch/record/1559262}{LHCb-CONF-2013-011}\relax
\mciteBstWouldAddEndPuncttrue
\mciteSetBstMidEndSepPunct{\mcitedefaultmidpunct}
{\mcitedefaultendpunct}{\mcitedefaultseppunct}\relax
\EndOfBibitem
\bibitem{Hsiao:2014tda}
Y.~K. Hsiao and C.~Q. Geng,
  \ifthenelse{\boolean{articletitles}}{\emph{{$f_J(2220)$ and hadronic \Bsb
  decays}}, }{}\href{http://dx.doi.org/10.1140/epjc/s10052-015-3317-9}{Eur.\
  Phys.\ J.\  \textbf{C75} (2015) 101},
  \href{http://arxiv.org/abs/1412.4900}{{\normalfont\ttfamily
  arXiv:1412.4900}}\relax
\mciteBstWouldAddEndPuncttrue
\mciteSetBstMidEndSepPunct{\mcitedefaultmidpunct}
{\mcitedefaultendpunct}{\mcitedefaultseppunct}\relax
\EndOfBibitem
\bibitem{PhysRevD.89.056003}
C.-K. Chua, \ifthenelse{\boolean{articletitles}}{\emph{Charmless two-body
  baryonic ${B}_{u,d,s}$ decays revisited},
  }{}\href{http://dx.doi.org/10.1103/PhysRevD.89.056003}{Phys.\ Rev.\
  \textbf{D89} (2014) 056003},
  \href{http://arxiv.org/abs/1312.2335}{{\normalfont\ttfamily
  arXiv:1312.2335}}\relax
\mciteBstWouldAddEndPuncttrue
\mciteSetBstMidEndSepPunct{\mcitedefaultmidpunct}
{\mcitedefaultendpunct}{\mcitedefaultseppunct}\relax
\EndOfBibitem
\end{mcitethebibliography}





 
\clearpage
\centerline{\large\bf LHCb collaboration}
\begin{flushleft}
\small
R.~Aaij$^{40}$,
B.~Adeva$^{39}$,
M.~Adinolfi$^{48}$,
Z.~Ajaltouni$^{5}$,
S.~Akar$^{59}$,
J.~Albrecht$^{10}$,
F.~Alessio$^{40}$,
M.~Alexander$^{53}$,
S.~Ali$^{43}$,
G.~Alkhazov$^{31}$
% ...

{\footnotesize \it
% ...
$ ^{5}$Clermont Universit{\'e}, Universit{\'e} Blaise Pascal, CNRS/IN2P3, LPC, Clermont-Ferrand, France\\
% ...
$ ^{10}$Fakult{\"a}t Physik, Technische Universit{\"a}t Dortmund, Dortmund, Germany\\
% ...
$ ^{31}$Petersburg Nuclear Physics Institute (PNPI), Gatchina, Russia\\
% ...
$ ^{39}$Universidad de Santiago de Compostela, Santiago de Compostela, Spain\\
$ ^{40}$European Organization for Nuclear Research (CERN), Geneva, Switzerland\\
% ...
$ ^{43}$Nikhef National Institute for Subatomic Physics, Amsterdam, The Netherlands\\
% ...
$ ^{48}$H.H. Wills Physics Laboratory, University of Bristol, Bristol, United Kingdom\\
% ...
$ ^{53}$School of Physics and Astronomy, University of Glasgow, Glasgow, United Kingdom\\
% ...
$ ^{59}$University of Cincinnati, Cincinnati, OH, United States\\
% ...
}
\end{flushleft}

\end{document}